\documentclass{aa}  

\usepackage{graphicx}
\usepackage{txfonts}
\usepackage{natbib} 
\bibpunct{(}{)}{;}{a}{}{,} 

\usepackage{verbatim} 

\usepackage{hyperref}   
\hypersetup{colorlinks=true,linkcolor=blue,citecolor=blue,filecolor=blue,urlcolor=blue}

\begin{document}

   \title{The role of stellar model input in correcting the asteroseismic scaling relations} 

   \subtitle{Red giant branch models}
   \author{G. Valle \inst{1, 2}\orcid{0000-0003-3010-5252}, M. Dell'Omodarme \inst{1}\orcid{0000-0001-6317-7872}, P.G. Prada Moroni
        \inst{1,2}\orcid{0000-0001-9712-9916}, S. Degl'Innocenti \inst{1,2}\orcid{0000-0001-9666-6066} 
}
\titlerunning{$f_{\Delta \nu}$}
\authorrunning{Valle, G. et al.}

\institute{
        Dipartimento di Fisica "Enrico Fermi'',
        Universit\`a di Pisa, Largo Pontecorvo 3, I-56127, Pisa, Italy\\
        \email{valle@df.unipi.it}
        \and
        INFN,
        Sezione di Pisa, Largo Pontecorvo 3, I-56127, Pisa, Italy
}

   \date{Received 10/02/2025; accepted  25/04/2025}

  \abstract
{}
{ This study investigates the variability of the theoretical correction factor, $f_{\Delta \nu}$, used in red giant branch (RGB) scaling relations, arising from different assumptions in stellar model computations. 
 }
{
Adopting a commonly used framework, we focused on a 1.0 $M_{\sun}$ star and systematically varied seven input parameters: the reference solar mixture, the initial helium abundance, the inclusion of microscopic diffusion and mass loss, the method for calculating atmospheric opacity, the mixing-length parameter, and the boundary conditions. Each parameter was tested using two distinct but physically plausible values to mimic possible choices of different evolutionary codes. For each resulting stellar model, we computed the oscillation frequencies along the RGB and derived the large frequency spacing, $\Delta \nu_0$. The correction factor $f_{\Delta \nu}$ was then calculated by comparing the derived $\Delta \nu_0$ with that predicted by the uncorrected scaling relations. 
}
{
We found substantial variability in $f_{\Delta \nu}$ across the different models. The variation ranged from approximately 1.3\% in the lower RGB to about 3\% at $\log g = 1.4$. This level of variability is significant, as it corresponds to roughly half the values typically quoted in the literature and leads to a systematic change in derived masses from 5\% to more than 10\%. The most significant contribution to this variability came from the choice of atmospheric opacity calculation (approximately 1.2\%), with a smaller contribution from the inclusion of microscopic diffusion (approximately 0.04\%). }
{ 
These results indicate that the choice of the reference stellar model has a non-negligible impact on the calculation of correction factors applied to RGB star scaling relations.
 } 

   \maketitle

\section{Introduction}\label{sec:intro}

Accurate determination of stellar fundamental parameters such as mass, radius, and age is crucial for understanding the evolutionary history of our Galaxy. Precision asteroseismology, enabled by satellite missions such as CoRoT \citep{Baglin2009}, {\it Kepler} \citep{Borucki2010}, K2 \citep{Howell2014}, and Transiting Exoplanet Survey Satellite  \citep[TESS;][]{Ricker2015},  has significantly advanced our ability to estimate these parameters.
This improvement stems from the ability to compare precise observational data with theoretical global asteroseismic parameters, namely the large frequency spacing, $\Delta \nu$, related to the acoustic radii of the stars \citep[e.g.][and references therein]{Dalsgaard2012}, and  the frequency of maximum oscillation power $\nu_{\rm max}$, related to the acoustic cut-off frequency of a star
\citep[see e.g.][]{Chaplin2008}. 
This approach leverages the asymptotic scaling relations \citep{Ulrich1986, Kjeldsen1995}
\begin{eqnarray}
\Delta \nu &\propto& \sqrt{\frac{M}{R^3}}\\
\nu_{\rm max} &\propto& \frac{M}{R^2 \sqrt{T_{\rm eff}}},
\end{eqnarray}
which are frequently used in inverted form to estimate stellar masses and radii from observed asteroseismic parameters \citep{Stello2009, Kallinger2010}:
\begin{eqnarray}
	\frac{M}{M_{\sun}} &=& 
	\left( \frac{\nu_{\rm max}}{f_{\nu_{\rm max}}  \nu_{\rm max, \sun}} \right)^3
	\left( \frac{\Delta \nu}{f_{\Delta \nu} \Delta \nu_{\sun}} \right)^{-4} 
	\left( \frac{T_{\rm eff}}{T_{\rm eff,\sun}} \right)^{3/2} \label{eq:Mc}\\
	\frac{R}{R_{\sun}} &=& 
	\left( \frac{\nu_{\rm max}}{f_{\nu_{\rm max}} \nu_{\rm max, \sun}} \right)
	\left( \frac{\Delta \nu}{f_{\Delta \nu} \Delta \nu_{\sun}} \right)^{-2} 
	\left( \frac{T_{\rm eff}}{T_{\rm eff,\sun}} \right)^{1/2},\label{eq:Rc}	
	\label{eq:scaling-corrected}
\end{eqnarray}
where $f_{\Delta \nu}$ and  $f_{\nu_{\rm max}}$ are correction factors. Since scaling relations were originally calibrated to the Sun, it is unsurprising that 
their validity in the red giant branch (RGB) phase has been questioned \citep[e.g.][]{Epstein2014b, Gaulme2016, Viani2017, Rodrigues2017, Zinn2019, Stello2022, Li2023}. Correction factors have been investigated by many authors across different ranges of mass, metallicity, and evolutionary phase \citep[e.g.][]{White2011, Sharma2016, Rodrigues2017, Guggenberger2017, Stello2022, Li2022, Pinsonneault2018}. 
At present, while it is possible to use stellar models and pulsation theory to investigate the correction factor $f_{\Delta \nu}$, understanding of the excitation, damping, and reflection of modes is still insufficient to do the same for $f_{\nu_{\rm max}}$, although some efforts exist in the literature \citep[e.g.][]{Viani2017, Zhou2020, Zhou2024}.
This issue is currently addressed by calibrating $f_{\nu_{\rm max}}$ to observations \citep[e.g.][]{Li2022, Pinsonneault2025}.

Concerning $f_{\Delta \nu}$, a widespread approach to correcting scaling relations is to obtain theoretically motivated corrections to the observed $\Delta \nu$,  to reconcile the observed
large frequency spacing  with the theoretical $\Delta \nu_{\rm s}$ assumed in
asymptotic pulsation theory \citep[e.g.][]{Sharma2016, Guggenberger2017, Stello2022}. These corrections depend on the evolutionary state of the star, as well as its mass, temperature, surface gravity, and metallicity. Their evaluation requires computing a large grid of stellar models, calculating oscillation frequencies using a stellar oscillation code, determining the average large frequency spacing $\Delta \nu_0$, and it with $\Delta \nu_{\rm s}$ from scaling relations.  In the following we adopt the definition
\begin{equation}
	f_{\Delta \nu} = \frac{\Delta \nu_0}{\Delta \nu_{\rm s}}.\label{eq:fdnu}
\end{equation}

Including correction factors, estimated masses and radii show general agreement for stars in eclipsing binary systems \citep[e.g.][and references therein]{Brogaard2018, Hekker2020}. 
Today, model-based corrections that account for stellar mass, evolutionary stage, and metallicity, such as those by \citet{Sharma2016}, are widely adopted in asteroseismic-based Galactic archaeology studies \citep[e.g.][]{Yu2018, Zinn2019, Warfield2021, Wang2023, Stasik2024, Warfield2024, Valle2024k2age}. 
A potential issue in adopting masses and radii from corrected scaling relations is the existence of discrepancies between theoretical and observed oscillation frequencies in stars with solar-like oscillations. One example is the surface effect, which originates from improper modelling of the stellar surface layers. Correcting for this effect remains an ongoing process \citep[][and references therein]{Ball2018, Jorgensen2019, Li2023}. 
Another concern is the possible discrepancy between observed and theoretical temperatures for stars on the RGB \citep{Martig2015, Valle2024k2age}. As shown by \citet{Gaulme2016}, a difference of 100 K in effective temperature corresponds to differences of about 3\% and 1\%, in estimated masses and radii, respectively.

Beyond these practical challenges, a fundamental theoretical concern that impacts the accuracy of the $f_{\Delta \nu}$ correction factor is the inherent flexibility that stellar modellers have in constructing the grids used to calculate these corrections. 
Specifically, the uncertainties in the input physics for stellar evolution codes introduce a significant degree of freedom in the selection of physical processes and parameters during model computation. Consequently, correction factors or calibrations derived from one set of stellar models cannot be reliably applied to models generated with different input physics.
The dependence of the correction factor on stellar models has been explored in several studies \citep[e.g.][]{Pinsonneault2018, Li2023, Pinsonneault2025}, revealing variations of a few percent due to differing model assumptions. However, these investigations overlook the intrinsic variability within the stellar models themselves.
Given the limited research on this topic, the primary objective of this article is to systematically investigate this issue using a test case.
We adopted the same methodological framework established by \citet{Sharma2016} but employed a different set of reference stellar models. In detail, we systematically varied seven key parameters, including initial chemical composition, evolutionary processes, and input physics, as outlined in Sect.~\ref{sec:method}. These variations were carefully chosen to represent a wide range of plausible choices in stellar modelling.
For each reference model, we computed the $f_{\Delta \nu}$  factors throughout the RGB evolution. By directly comparing the results obtained under different reference scenarios and performing a statistical analysis of the impact of various parameters, we aim to shed light on the influence of stellar model uncertainties on the $f_{\Delta \nu}$ corrections.

\section{Methods}\label{sec:method}
\subsection{Stellar models}

Stellar models were computed with Modules for Experiments in Stellar Astrophysics (MESA) r24.03.01 for a 1.0 $M_{\sun}$ star.  The models evolved from the pre-main sequence up to the RGB until $\log g = 1.4$ dex. 
These values were chosen to represent a typical star in the APO-K2 catalogue \citep{Stasik2024}, covering the RGB range spanned by the catalogue stars. Seven key parameters that impact stellar evolution and asteroseismic observables were varied dichotomously to represent different legitimate choices made by stellar modellers (see Table~\ref{tab:param}). 
Given the current theoretical and observational uncertainties, none of the options considered can be definitively considered more accurate or appropriate. All explored configurations were implemented using the standard MESA distribution, without the development of custom code.

\begin{table}
	\caption{Parameters varied in the stellar models computation and their corresponding values.}\label{tab:param}
	\centering
	\begin{tabular}{lccc}
		\hline\hline
		Parameter & Label & Option 1 & Option 2\\
		\hline
		Solar mixture & p1 & GS98 & AP09 \\
		$\Delta Y/\Delta Z$ & p2 & 2.0 & 1.5 \\
		Diffusion & p3 & off & on\\
		Mass loss $\eta$ & p4 & 0.0 & 0.2 \\
		Atmosphere $\kappa$ & p5 & fixed & varying \\
		$\alpha_{\rm ml}$ & p6 & solar & solar - 0.1 \\
		Boundary conditions & p7 & Hopf & Eddington \\
		\hline  
	\end{tabular}
\end{table}

The first group of parameters concerned the initial chemical abundances. This included the reference solar mixture and the initial helium abundance.
For the former, we adopted two widely used mixtures: the \citet{GS98} mixture (hereafter GS98) and the \citet{AGSS09} mixture (hereafter AP09).
For the latter, we set the initial helium abundance as $Y = Y_p +  \Delta Y/\Delta Z$, where $Y_p = 0.2471$ from \citet{Planck2020}.  The helium-to-metal enrichment ratio,  $\Delta Y/\Delta Z$, was set to 2.0 and 1.5, values consistent with the current uncertainty in this parameter \citep[e.g.][and references therein]{Tognelli2021}. 
The initial metallicity was fixed at [Fe/H] = -0.5, to match the median metallicity in the APO-K2 sample. This corresponds to $Z = 0.00538$ and 0.00540 for $\Delta Y/\Delta Z = 2.0$ and 1.5, respectively, when adopting the GS98 solar mixture. The same values of $Z$ were also adopted for the AP09 solar mixture. 
Stellar data fitting often uses the surface [Fe/H] to represent metallicity, and the correspondence between [Fe/H] and the total metallicity $Z$ depends on the chosen reference solar mixture. Since a change of $Z$ has a significant effect on stellar evolution, fixing [Fe/H] for model comparisons could introduce spurious differences. Therefore, we preferred to fix $Z$  rather than [Fe/H] to avoid inflating the differences due to the mixture change.

The second group of parameters represented the inclusion of additional physical processes, specifically microscopic diffusion and mass loss. For microscopic diffusion, two extreme scenarios were considered.  While microscopic diffusion has been shown to be efficient in the Sun \citep[see e.g.][]{Bahcall2001},  its efficiency in Galactic globular cluster stars remains the subject of debate \citep[see e.g.][]{Korn2007, Gratton2011, Nordlander2012,Gruyters2014}. Therefore, we computed a set of models accounting for gravitational settling, without radiative acceleration, following the standard MESA implementation \citep{MESA2018}, and a set excluding it.
For mass loss, the \citet{Reimers1975} mass-loss formalism was adopted, with 
the  $\eta$  parameter ranging from 0.0 to 0.2. These values are consistent with those found in studies of open clusters \citep{Miglio2012MNRAS, Handberg2017}.  

The remaining parameters were related to the algorithmic computation of the stellar models. The opacity, $\kappa$ , in the atmosphere was assumed to be either uniform throughout the atmosphere, with a value set by the outermost cell, or varied with depth using the MESA {\it kap} module at the local optical depth $\tau$ \citep{Jermyn2023}. 
The mixing-length parameter, $\alpha_{\rm ml}$, was fixed at the solar-calibrated value or decreased by 0.1 to model potential variations throughout stellar evolution \citep{Magic2014}.
A solar calibration was performed for each set of model parameters. This was achieved by iteratively refining the metallicity $Z$, initial helium abundance $Y$, and mixing-length parameter using a Levenberg-Marquardt algorithm\footnote{Implemented via the \textit{marqLevAlg} R library \citep{R, marqLevAlg}.}. The calibration aimed to reproduce the observed solar radius, luminosity, logarithm of the effective temperature, and surface $Z/X$ at the Sun's current age. The iterative process terminated when the sum of the squared relative differences between the model outputs and the target values was less than $10^{-7}$. Mass loss was not included in this calibration process.

Finally, boundary conditions were set using either the solar-calibrated Hopf 
$T(\tau)$ relation \citep{MESA2013}, a ﬁt to Model C of the solar atmosphere by \citet{Vernazza1981}, or the solar-calibrated Eddington $T(\tau)$ relation.
While this parameter set does not encompass the full range of variability affecting models and pulsation frequencies, it includes many key factors that undoubtedly impact RGB evolution\footnote{The MESA inlists used in the computations are available at \url{https://github.com/mattdell71/fdnu}.}.

\subsection{Experiment design}

\begin{table}
	\caption{Design matrix of the 32 orthogonal combinations of the seven considered parameters.}\label{tab:design}
	\centering
	\begin{tabular}{lccccccc}
		\hline\hline
		& p1 & p2 & p3 & p4 & p5 & p6 & p7 \\ 
		\hline
		1 & 1 & 0 & 1 & 0 & 1 & 1 & 1 \\ 
		2 & 1 & 1 & 0 & 1 & 0 & 0 & 0 \\ 
		3 & 1 & 1 & 0 & 1 & 1 & 0 & 1 \\ 
		4 & 1 & 0 & 0 & 1 & 0 & 1 & 1 \\ 
		5 & 0 & 1 & 1 & 0 & 1 & 0 & 1 \\ 
		6 & 1 & 1 & 1 & 1 & 1 & 1 & 0 \\ 
		7 & 1 & 0 & 0 & 1 & 1 & 0 & 0 \\ 
		8 & 0 & 0 & 0 & 0 & 0 & 0 & 1 \\ 
		9 & 0 & 1 & 1 & 1 & 0 & 1 & 1 \\ 
		10 & 0 & 0 & 1 & 0 & 0 & 1 & 1 \\ 
		11 & 0 & 0 & 0 & 1 & 1 & 1 & 1 \\ 
		12 & 0 & 1 & 0 & 1 & 0 & 0 & 1 \\ 
		13 & 1 & 0 & 1 & 1 & 0 & 0 & 1 \\ 
		14 & 1 & 0 & 1 & 1 & 0 & 1 & 0 \\ 
		15 & 0 & 1 & 0 & 1 & 0 & 1 & 0 \\ 
		16 & 0 & 1 & 0 & 0 & 1 & 0 & 0 \\ 
		17 & 1 & 1 & 1 & 0 & 0 & 0 & 1 \\ 
		18 & 0 & 1 & 1 & 1 & 1 & 0 & 0 \\ 
		19 & 1 & 1 & 0 & 0 & 1 & 1 & 0 \\ 
		20 & 0 & 0 & 0 & 0 & 0 & 0 & 0 \\ 
		21 & 0 & 1 & 0 & 0 & 1 & 1 & 1 \\ 
		22 & 1 & 1 & 1 & 1 & 1 & 1 & 1 \\ 
		23 & 0 & 0 & 1 & 1 & 1 & 0 & 1 \\ 
		24 & 1 & 0 & 1 & 0 & 1 & 0 & 0 \\ 
		25 & 1 & 0 & 0 & 0 & 1 & 0 & 1 \\ 
		26 & 0 & 0 & 1 & 0 & 1 & 1 & 0 \\ 
		27 & 1 & 0 & 0 & 0 & 0 & 1 & 0 \\ 
		28 & 0 & 0 & 0 & 1 & 1 & 1 & 0 \\ 
		29 & 1 & 1 & 0 & 0 & 0 & 1 & 1 \\ 
		30 & 0 & 0 & 1 & 1 & 0 & 0 & 0 \\ 
		31 & 0 & 1 & 1 & 0 & 0 & 1 & 0 \\ 
		32 & 1 & 1 & 1 & 0 & 0 & 0 & 0 \\ 
		\hline
	\end{tabular}
\end{table}

To assess the importance of individual parameters, a full factorial design involving evaluation of all $2^7$ combinations was deemed impractical due to inefficiency. In a computer experiment, where random variability between replicates is absent, an orthogonal array design -- a generalisation of the Latin hypercube design -- was more appropriate due to its higher efficiency \citep{Art1992, Hedayat1999}.    
In an orthogonal array, each factor (i.e. each parameter in Table ~\ref{tab:param})  appears with equal frequency at each level (i.e. each option in Table~\ref{tab:param}), and each pair of factors is included with the same frequency for each pair of levels.
Therefore, in our experimental design, each pair of factors was considered an equal number of times for the four possible combinations of options.
Since our primary goal was to assess the impact of individual parameters (i.e. main effects), rather than their interactions (which were assumed to be of minor relevance), an orthogonal design was appropriate.
   
Using the R library {\it DoE.base} \citep{DoE}, we generated a set of 32 random orthogonal combinations of the seven parameters. The design matrix used in the experiment is presented in Table~\ref{tab:design}, where each factor is coded as 0 or 1, with 0 indicating that Option 1 in Table~\ref{tab:param} was adopted.

\subsection{Oscillation frequency computation}   

The oscillation frequencies were calculated using GYRE 7.1 \citep{Townsend2013, Townsend2018} from stellar structure profiles computed with MESA.  The structure of the stellar atmosphere was included in the stellar profiles.  
For each stellar track, oscillation frequencies were computed from $\log g = 3.2$  to $\log g = 1.4$ in steps of 0.01, following the RGB evolution. We adopted the approach of \citet{White2011} to derive $\Delta \nu_0$ for each model, mimicking how this parameter is measured from the data \citep{Sharma2016, Stello2022}. 
The observed oscillation power spectrum of solar-like stars is characterised by a typical Gaussian-like envelope, peaked at $\nu_{\rm max}$. To mimic the observable frequency ranges for each model, the frequencies were weighted based on their position within a Gaussian envelope centred at $\nu_{\rm max}$, with a full width at half maximum given by $0.66 \, \nu_{\rm max}^{0.88}$ \citep{Mosser2012, Bellinger2016}. 
The frequency of maximum oscillation power was derived using the scaling relation, adopting the MESA default $\nu_{\rm max, \sun} = 3078$ $\mu$Hz from \citet{Lund2017}.
The value of $\Delta \nu_0$  was then obtained by weighted regression of the oscillation frequencies versus their radial order.

Corresponding values of  $\Delta \nu_s$ were obtained from the scaling relation, with MESA default $\Delta \nu_{\sun} = 134.91$ $\mu$Hz from \citet{Lund2017}. The ratio $f_{\Delta \nu}$ of the computed to scaling relation large frequency separation  was then computed for each model, as described in Eq.~(\ref{eq:fdnu}).

\section{Results}\label{sec:results}

\begin{figure}
	\centering
	\resizebox{\hsize}{!}{\includegraphics{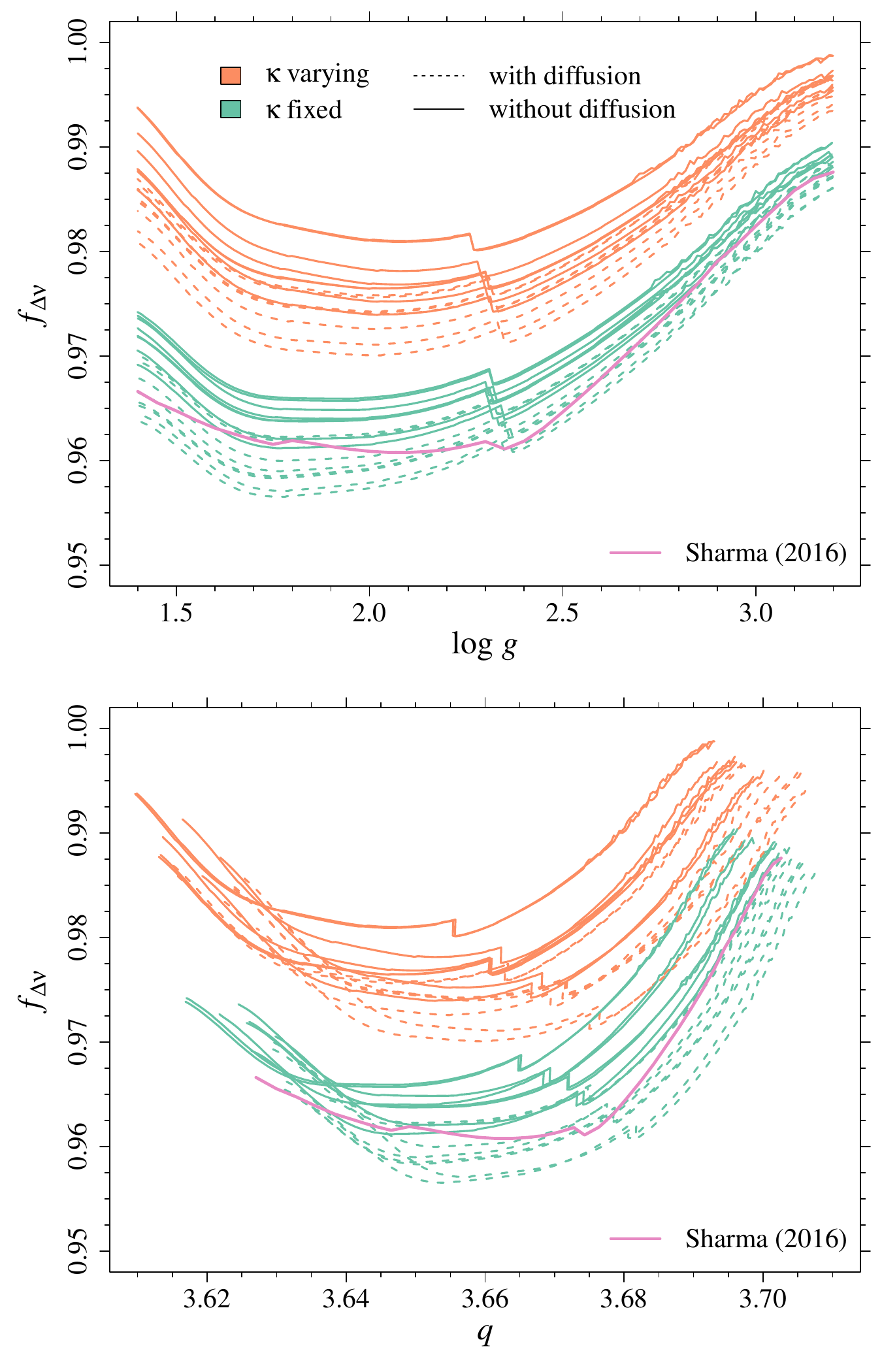}}
	\caption{
 Evolution of the $f_{\Delta \nu}$ ratio  for the 32 parameter combinations considered.
 Top panel: Trend of $f_{\Delta \nu}$ vs $\log g$.
The solid violet line indicates the reference model with $M=1.0$ $M_{\sun}$, $Z=0.00477$ , based on the \citet{Sharma2016} data set. Different colours represent models with varying (orange) and fixed (green) opacity $\kappa$ in the atmosphere, the primary source of systematic uncertainty in $f_{\Delta \nu}$. Line styles differentiate models calculated with (dashed) and without (solid) microscopic diffusion, which is the second most significant contributor to systematic uncertainty in $f_{\Delta \nu}$.
Bottom panel: Same as the top panel, but shown as a function of the $q$ parameter.
}
	\label{fig:fdnu}
\end{figure}

The trends of $f_{\Delta \nu}$ versus $\log g$  are shown in Fig.~\ref{fig:fdnu} for all 32 cases considered. The trend from \citet{Sharma2016} for $Z=0.00477$ is also included as a reference\footnote{This is the $Z$ value closest to our choice of $Z=0.00538$. Minor differences arose considering $Z= 0.00601$, the next value available in the \citet{Sharma2016} grid.}. The qualitative behaviour is consistent across all 32 considered cases and agrees with the results of \citet{Sharma2016} results.
The most notable feature is the significant variability observed in $f_{\Delta \nu}$.
The maximum difference in the correction factor increases almost steadily from 0.013 to 0.030 as $\log g$ decreases from 3.2 to 1.4. Given the fourth-power dependence of the asteroseismic mass on $f_{\Delta \nu}$, this variability leads to systematic differences in the masses estimated from the corrected scaling relations, ranging from about 5\% to more than 10\%. 
This range of difference in the correction factor depends only slightly on the stellar mass. In Appendix~\ref{app:mass} we present the results for models with masses of 0.8 $M_{\sun}$ and 1.2 $M_{\sun}$, respectively. Their analysis confirms the robustness of the results presented here.

In practical applications of the $f_{\Delta \nu}$ correction, selecting the appropriate factor from a computed grid requires consideration of not only $\log g$, but also effective temperature, mass, and metallicity. For instance, \citet{Sharma2016} and \citet{Stello2022} utilise an interpolating variable, $q$, that encapsulates the combined influence of effective temperature and $\log g$:
\begin{equation}
q = \log T_{\rm eff} + 0.05 \, \log g \, \left( \tanh \frac{\log g - 4.5}{0.25} +1 \right).
\end{equation}
The bottom panel of Fig.~\ref{fig:fdnu} shows that the tracks cross at $q \leq 3.64$, which corresponds to the terminal part of the RGB. As in the previous analysis, we found that the maximum difference in the correction factor increases almost steadily from 0.015 to 0.032 as $q$ increases from 3.62 to 3.69.  
Since the tested cases represent reasonable and justifiable choices in stellar model computation, this variability poses a theoretical challenge when applying corrections to the scaling relations.

\begin{table}
	\caption{Coefficients and statistical significance from the linear model describing the dependence of $f_{\Delta \nu}$ on stellar model parameters. }\label{tab:lm}
	\centering
	\begin{tabular}{lrrrr}
		\hline
		& Estimate & Error & $t$ value & $P$ value \\ 
		\hline
		$\beta_0$ & $9.861 \cdot 10^{-1}$ & $2.74 \cdot10^{-4}$ & 3605 & $<  10^{-16}$ \\ 
		$\beta_1$ & $-9.237 \cdot 10^{-4}$ & $3.99 \cdot 10^{-5}$ & -23.2 & $<  10^{-16}$ \\ 
		$\beta_2$ & $-1.179 \cdot 10^{-5}$ & $3.99 \cdot 10^{-5}$ & -0.3 & 0.77 \\ 
		$\beta_3$ & $-3.594 \cdot 10^{-3}$ & $3.99 \cdot 10^{-5}$ & -90.1 & $<  10^{-16}$ \\ 
		$\beta_4$ & $-1.947 \cdot 10^{-5}$ & $3.99 \cdot 10^{-5}$ & -0.5 & 0.63 \\ 
		$\beta_5$ & $1.216 \cdot 10^{-2}$ & $3.99 \cdot 10^{-5}$ & 305 & $<  10^{-16}$ \\ 	
		$\beta_6$ & $2.423 \cdot 10^{-3}$ & $3.99 \cdot 10^{-5}$ & 60.7 & $<  10^{-16}$ \\ 
		$\beta_7$ & $2.033 \cdot 10^{-3}$ & $3.99 \cdot 10^{-5}$ & 50.9 & $<  10^{-16}$ \\ 
		\hline
	\end{tabular}
	\tablefoot{Columns 1 and 2: least-squares estimates of regression
		coefficients and their errors; column 3: $t$-statistic testing the
		statistical significance of the covariates; column 4: the $P$ value for the significance test.}
\end{table}

To investigate the origin of this variability, a linear model was fitted to describe the dependence of the correction factor on the stellar model parameters. To account for the non-linear dependence on $\log g$, a dummy factor was included in the model to adjust for differences in the mean values of $f_{\Delta \nu}$ across the RGB points for each case. However, we found that excluding this stratification factor had a negligible impact on the estimated model parameters.
The final model is given by:
\begin{equation}
f_{\Delta \nu} = \beta_0 + \sum_{i=1}^7 \beta_i p_i,   
\end{equation}
where $\beta$ denoted the regression coefficients and $p_i$ are dichotomous categorical variables representing the two possible values of the $i$-th parameter in Table ~\ref{tab:param}.
The model coefficients are presented in Table ~\ref{tab:lm}. The most significant source of variability in the correction factor was the computation of opacity $\kappa$ in the atmosphere. Using a variable $\kappa$ in the stellar model computations increases $f_{\Delta \nu}$ by approximately 1.2\% which is approximately one third to one half of the \citet{Sharma2016} correction.  
The second most important factor was the inclusion of microscopic diffusion, as stellar models that accounted for this process yielded a lower $f_{\Delta \nu}$ by about 0.4\%. 
The other parameters had a lower impact, around 0.1\% to 0.2\%, with the exceptions of the helium-to-metal enrichment ratio and the inclusion of mass loss, which were found to be statistically insignificant. However, the importance of the helium-to-metal enrichment ratio may increase at higher initial metallicities due to the assumed link between $Z$ and initial $Y$.

\section{Conclusions}\label{sec:conclusions}

We investigated the variability in the computation of the correction factor $f_{\Delta \nu}$ for the scaling relations. We adopted a framework identical to that of  \citet{Sharma2016}, which is widely used in the literature. For the test case of $M = 1.0$ $M_{\sun}$, different stellar models were computed, differing in the choice of  seven input parameters: the reference solar mixture, the initial helium abundance, the inclusion of microscopic diffusion, the inclusion of mass loss, the computation of the opacity $\kappa$ in the atmosphere, the adopted mixing-length parameter value, and the boundary conditions. Each parameter was varied between two different options, each of them being a legitimate choice for stellar model computations.

By adopting an efficient orthogonal array design, we selected 32 combinations of these parameters, and for each, stellar tracks were evolved with MESA r24.03.01 up to $\log g = 1.4$ in the RGB phase. For each track, the oscillation frequencies were calculated in the RGB phase with GYRE 7.1, and the large spacing $\Delta \nu_0$ was computed as in \citet{White2011}. Finally, the correction factor $f_{\Delta \nu}$ was calculated as the ratio of $f_{\Delta \nu}$ to the value of the uncorrected scaling relations.  

The correction factor showed substantial variability across the 32 selected reference stellar models. The range of variability in $f_{\Delta \nu}$ was lower, approximately 1. 3\% in the lower RGB, and steadily increased to approximately 3\% in $\log g = 1.4$. This variability is significant, being close to corrections frequently adopted in the literature, such as that of \citet{Sharma2016}, and would cause variations in corrected mass from 5\% to 10\% in the absence of recalibration, such as that discussed by \citet{Pinsonneault2025}.
The most important contribution to this variability was the modelling of the opacity $\kappa$ in the stellar atmosphere (about 1.2\%), followed by the inclusion of microscopic diffusion (about 0.4\%). 

The most recent Galactic archaeology studies \citep[e.g.][]{Zinn2019, Warfield2021, Wang2023, Stasik2024, Valle2024k2age} adopt corrections similar to those discussed here.
In this respect, the relevance of the variability in the correction factor is clear, because the random variability affecting mass estimates directly translates into variability in the derived stellar ages. In fact, the uncertainty that affects determinations of the effective temperature of RGB stars, both theoretical \citep[e.g.][]{Straniero1991, Salaris1996, Vandenberg1996}  and observational \citep[e.g.][]{Hegedus2023, Yu2023}, has raised concerns about the direct use of this constraint in grid-based fitting \citep{Martig2015, Warfield2021, Vincenzo2021}. To address this issue, stellar model grids for RGB stars are usually interpolated using the asteroseismic mass as a constraint \citep{Martig2015, Warfield2021, Warfield2024} or the theoretical effective temperature corresponding to that mass and $\Delta \nu$ \citep{Valle2024k2age}. Given the dominant role of stellar mass in determining RGB stellar ages, the systematic uncertainty reported here raises questions about the accuracy attainable when estimating the age of RGB stars.  

Although the model parameters tested are likely among the main sources of variability in the calculation of $f_{\Delta \nu}$, they are not the only factors. Moreover, their influence may vary across different metallicity ranges and stellar masses. For example, in more massive stars, microscopic diffusion is expected to play a minor role due to shorter evolutionary timescales. Conversely, for these stars, convective core overshooting is expected to impact the systematic variability of the correction factor. 
Despite these limitations, the findings of this study highlight the importance of considering systematic variability in correction factors for scaling relations that rely on a stellar model reference scenario.

\begin{acknowledgements}
G.V., P.G.P.M. and S.D. acknowledge INFN (Iniziativa specifica TAsP) and support from PRIN MIUR2022 Progetto "CHRONOS" (PI: S. Cassisi) finanziato dall'Unione Europea - Next Generation EU.
\end{acknowledgements}

\bibliographystyle{aa}
\bibliography{biblio}

\appendix

\section{Mass dependence}\label{app:mass}

\begin{figure}[h!]
	\centering
	\resizebox{\hsize}{!}{\includegraphics{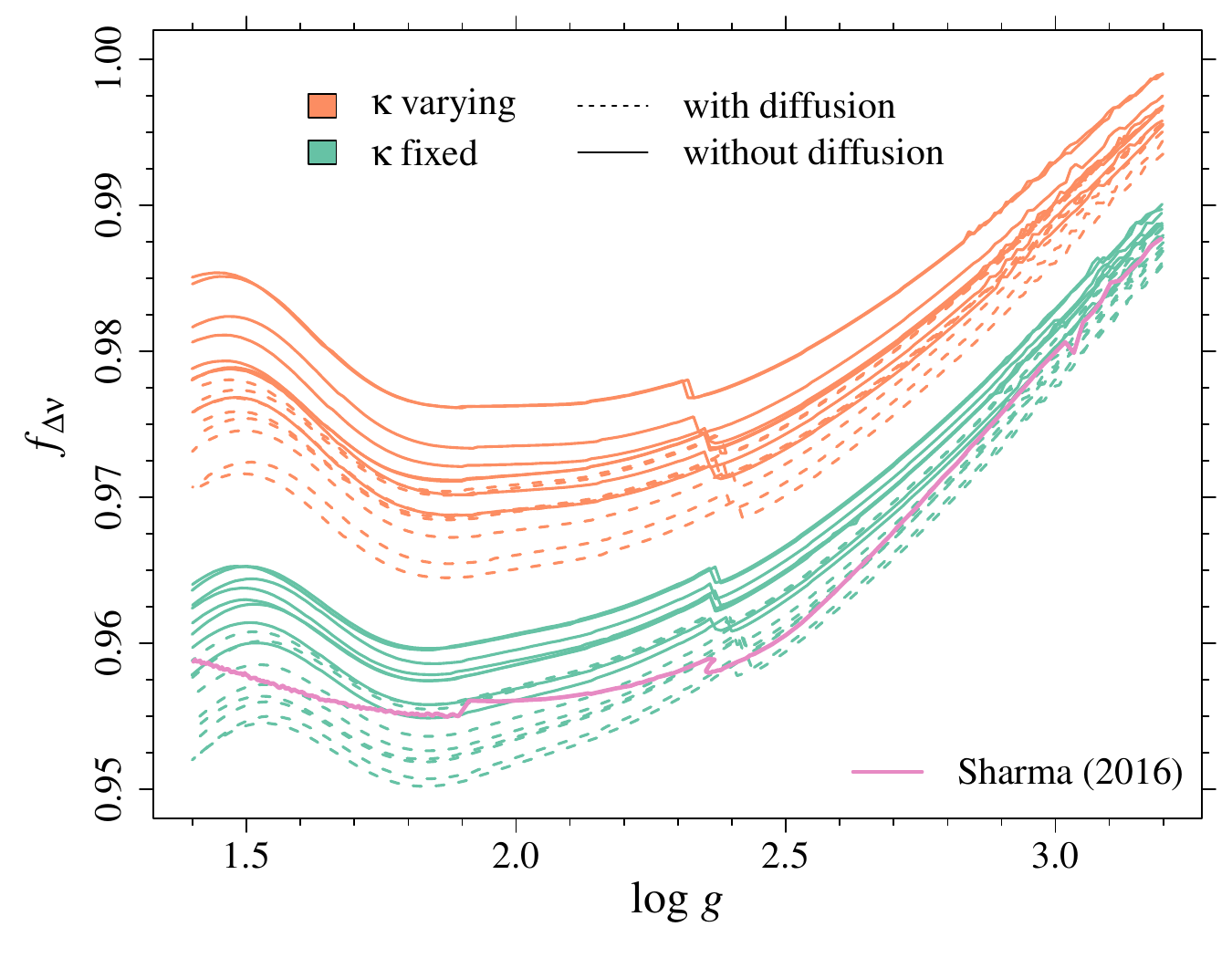}}\\
    \resizebox{\hsize}{!}{\includegraphics{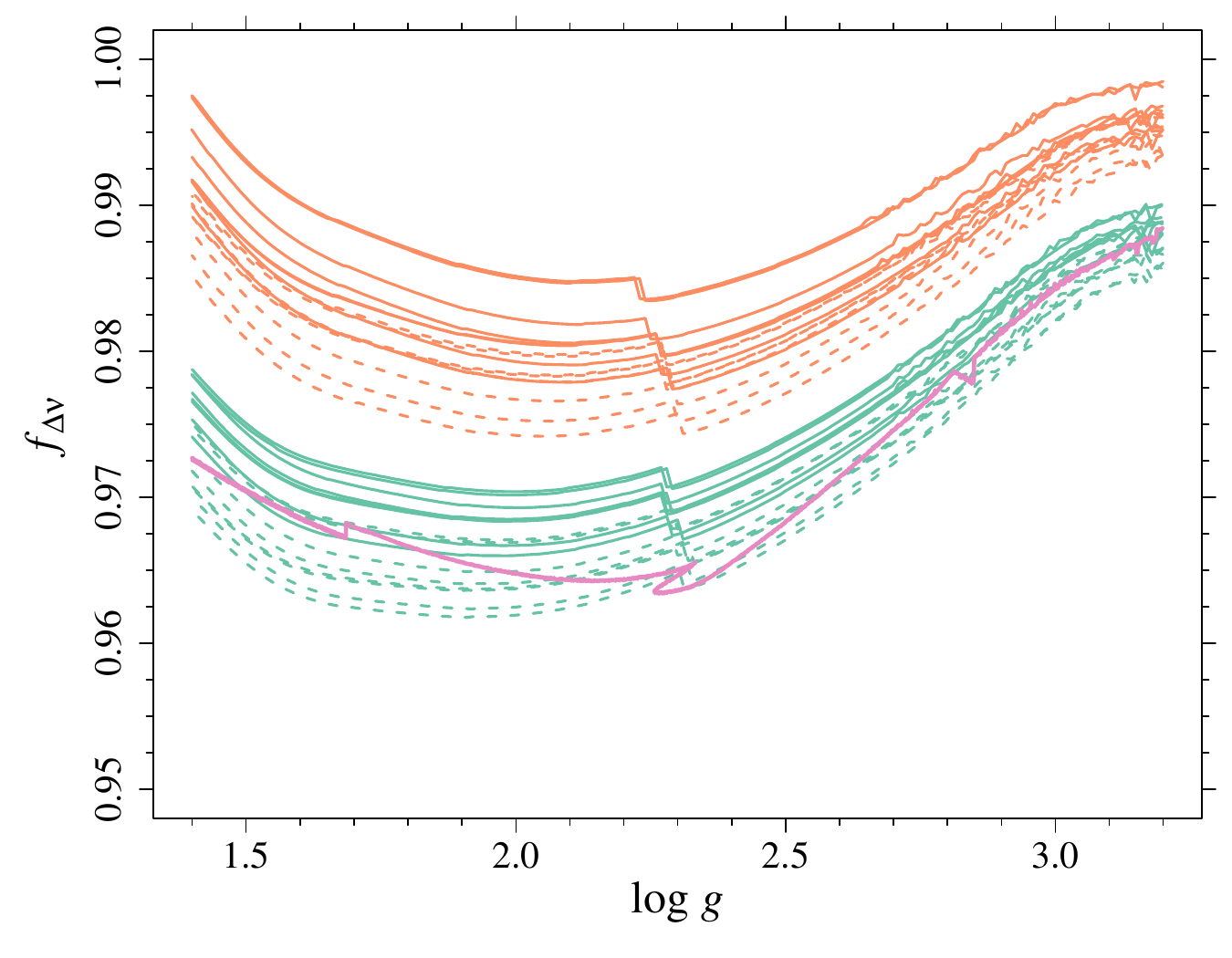}}
	\caption{
 Evolution of the $f_{\Delta \nu}$ ratio with $\log g$ for the 32 parameter combinations considered.
 ({\it Top}): models computed with $M = 0.8$ $M_{\sun}$. The colour and line codes are the same as in Fig.~\ref{fig:fdnu}.
  ({\it Bottom}): same as in the top panel, but for $M = 1.2$ $M_{\sun}$. 
 }
	\label{fig:fdnu-masses}
\end{figure}

This section compares the correction factors $f_{\Delta \nu}$ for stellar models with masses 0.8 $M_{\sun}$ and 1.2 $M_{\sun}$. As illustrated in Fig.~\ref{fig:fdnu-masses}, the variation in $f_{\Delta \nu}$ due to input physics is nearly identical to that observed for the 1.0 $M_{\sun}$ model, discussed in the main text.

\end{document}